\documentclass[twocolumn,aps,pra,superscriptaddress,nofootinbib,notitlepage,showpacs,floatfix,longbibliography]{revtex4-1}

\usepackage{amsmath,amssymb,amsthm,dsfont}
\usepackage{braket}
\usepackage{empheq}
\usepackage{subfigure}
\usepackage{cancel}
\usepackage[pdfstartview=FitH]{hyperref}
\hypersetup{
    colorlinks=true,       
    linkcolor=blue,          
    citecolor=magenta,        
    filecolor=magenta,      
    urlcolor=blue,           
    runcolor= blue
}

\usepackage{url}
\usepackage{color}
\usepackage{graphicx}
\usepackage{amsthm}

\DeclareMathOperator{\Tr}{\operatorname{Tr}}

\begin{document}

\title{Non-Markovianity of the Post Markovian Master Equation}

\author{Chris Sutherland}
\affiliation{Department of Physics and Astronomy, University of Southern California, Los Angeles, California 90089, USA}
\affiliation{Center for Quantum Information Science \& Technology, University of Southern California, Los Angeles, California 90089, USA}
\author{Todd A. Brun}
\affiliation{Department of Physics and Astronomy, University of Southern California, Los Angeles, California 90089, USA}
\affiliation{Center for Quantum Information Science \& Technology, University of Southern California, Los Angeles, California 90089, USA}
\affiliation{Department of Electrical Engineering, University of Southern California, Los Angeles, California 90089, USA}
\author{Daniel A. Lidar}
\affiliation{Department of Physics and Astronomy, University of Southern California, Los Angeles, California 90089, USA}
\affiliation{Center for Quantum Information Science \& Technology, University of Southern California, Los Angeles, California 90089, USA}
\affiliation{Department of Electrical Engineering, University of Southern California, Los Angeles, California 90089, USA}
\affiliation{Department of Chemistry, University of Southern California, Los Angeles, California 90089, USA}

\date{\today}

\begin{abstract}
An easily solvable quantum master equation has long been sought that takes into account memory effects induced on the system by the bath, i.e., non-Markovian effects. We briefly review the Post-Markovian master equation (PMME), which is relatively easy to solve, and analyze a simple example where solutions obtained exhibit non-Markovianity. We apply the distinguishability measure introduced by Breuer et al., and we also explicitly analyze the divisibility of the associated quantum dynamical maps. We give a mathematical condition on the memory kernel used in the PMME that guarantees non-CP-divisible dynamics. 
\end{abstract}


\maketitle
\section{Introduction}\label{intro}
The study of open quantum systems leads to a theoretically rich and experimentally useful theory~\cite{breuer2002theory,alicki_quantum_2007}. It allows us to make concrete predictions about a quantum system of interest that is interacting with its environment. One of the most commonly used equations to model open quantum system dynamics is the Gorini-Kossakowski-Sudarshan-Lindblad (GKSL) master equation~\cite{gorini1976completely,Lindblad:76}, mainly due to its easily solvable nature and the fact that it satisfies the condition of complete positivity. One of the main assumptions that goes into its derivation, and that makes it easily solvable is the assumption of Markovianity or being `memoryless'. This means that although the quantum system is interacting with a bath, no information about past states of the system flows back from the bath; the bath `forgets' about earlier states of the system in a very short time. That is, the evolution of a quantum system at time $t$ depends only on its density matrix $\rho(t)$ and not on its state $\rho(t')$ at earlier times $t'<t$.

Unfortunately, although the assumption of Markovianity allows for a pleasing simplification, the Lindblad master equation is only an approximation, and non-Markovian effects are often too important to neglect. At the other extreme, the formally exact Nakajima-Zwanzig master equation~\cite{zwanzig1960ensemble} is too hard to solve. Hence compromises leading to master equations that are both easily solvable and account for non-Markovian effects are desirable. This is particularly true in light of recent developments in quantum computation and quantum annealing, where non-Markovian effects often play an important role~\cite{Smirnov:2018aa}. There has already been much work on this problem, e.g., Gaussian~\cite{ferialdi2016exact}, quantum collisional models \cite{budini2013embedding}, and time-convolutionless master equations~\cite{smirne2010nakajima}. 
Here we focus on the post-Markovian master equation (PMME)~\cite{shabani2005completely}. The PMME was derived via an interpolation between the generalized measurement interpretation of the exact Kraus operator sum map~\cite{kraus1983states} and the continuous measurement interpretation of Markovian-limit dynamics~\cite{breuer2004genuine}. Previous work implied that the PMME was essentially Markovian~\cite{mazzola2010phenomenological}. This claim was subsequently countered in~\cite{budini2014post}. Our goal in this work is to revisit the question of the (non-)Markovianity of the PMME. We confirm that the PMME can describe non-Markovian dynamics, and provide a 
simple example to illustrate this.

The structure of this paper is as follows. 
In Section~\ref{pmmenonmarkovmeasure} we outline the definitions and measures of non-Markovianity that we use here to study the PMME. We describe what it means for quantum dynamics to be non-CP-divisible and how that relates to quantum non-Markovianity, and explain why an increase of distinguishability between two distinct initial quantum states is a witness of non-Markovianity~\cite{breuer2009measure}. We also briefly explain how the PMME is derived. In~Section \ref{markov}, we briefly summarize the reasoning behind the work which stated that the PMME is essentially Markovian~\cite{mazzola2010phenomenological}, and the more recent work offering evidence to the contrary~\cite{budini2014post}. Then, in Section~\ref{nonmarkov}, we describe the simple physical scenario of a qubit dephasing with a bath and show how the PMME accounts for non-Markovian effects that are not captured by the Lindblad equation. Finally, in Section \ref{Cpkerncondition} we give a mathematical condition on the memory kernel used in the PMME which guarantees non-CP-divisible dynamics. Supporting information for Section  \ref{markov} is given in the Appendix.

\section{Quantum non-Markovianity and the Post Markovian master equation}\label{pmmenonmarkovmeasure}
\subsection{Quantum non-Markovianity}
The problem of quantifying and describing quantum non-Markovianity has been the subject of deep study in recent years. Several key measures, witnesses, and definitions of quantum non-Markovianity are now well accepted~\cite{rivas2014quantum,Chruscinski:2017aa,wakakuwa2017operational}. Here we give a brief summary of the approach contained in~\cite{breuer2016colloquium}. Quite generally, quantum dynamics is described by quantum dynamical maps:
\begin{equation}\label{generalme}
\rho_{t}=\Phi_{t}(\rho_{0}),
\end{equation}
where $\Phi_{t}$ is a completely-positive trace-preserving (CPTP), time dependent map with $\Phi_{0}=I$ (for a more general class see~\cite{Dominy:14,Dominy:2016xy}).  The Markovian quantum master equation is a special case of this, where the quantum dynamical map (superoperator) $\Phi_{t}$ can be written as
\begin{equation}
\Phi_{t}=T\text{exp}\Bigl[\int_{0}^{t}d\tau\mathcal{L}_{\tau}\Bigr],
\end{equation}
where the $T$ denotes time ordering and $\mathcal{L}_{\tau}$ is the GKSL generator~\cite{gorini1976completely,Lindblad:76}, 
\begin{equation}\label{markovlindblad}
\mathcal{L}_{t}\rho=-i[H,\rho]+\sum_{k}\gamma_{k}\left(L_{k}\rho L_{k}^{\dagger}-\frac{1}{2}\{L_{k}^{\dagger}L_{k},\rho\}\right).
\end{equation}
Therefore, Eq.~\eqref{generalme} becomes 
\begin{equation}
\dot{\rho}=\mathcal{L}_{t}\rho.
\end{equation}
For this process to represent a completely positive one-parameter semigroup, the coefficients must satisfy $\gamma_{k}\geq 0$. This is known as the GKSL theorem.%
\footnote{This is actually a pair of theorems that were discovered independently and nearly simultaneously, for the finite dimensional case by Gorini, Kossakowski, and Sudarshan~\cite{gorini1976completely}, and the infinite dimensional case by Lindblad~\cite{Lindblad:76}; for a detailed account of this history see~\cite{Chruscinski:2017ab}.} 

An interesting class of dynamical maps $\Phi_{t}$ are those for which the inverse process $\Phi^{-1}_t$ exists. Then for $t_{2}\geq t_{1} \geq 0$, one can define a two-parameter family of maps given by
\begin{equation}
\Phi_{t_{2},t_{1}}=\Phi_{t_{2}}\Phi_{t_{1}}^{-1},
\end{equation}
such that $\Phi_{t_{2},0}=\Phi_{t_{2}}$ and
\begin{equation}\label{divisibility}
\Phi_{t_{2},0}=\Phi_{t_{2},t_{1}}\Phi_{t_{1},0}.
\end{equation}
In this case one can always write down a time-local quantum master equation with the following form:
\begin{align}\label{eq:generalMEform}
\dot{\rho}&=\mathcal{K}_{t}\rho=-i[H(t),\rho] \nonumber \\
&+\sum_{k}\gamma_{k}(t)\bigg(L_{k}(t)\rho L_{k}^{\dagger}(t)-\frac{1}{2}\{L_{k}^{\dagger}(t)L_{k}(t),\rho\}\bigg). 
\end{align}
Note the explicit time dependence in the Hamiltonian, Lindblad operators, and the rates. The $\gamma_{k}(t)$ coefficients in Eq.~\eqref{eq:generalMEform} need not be positive. 

The two-parameter family of quantum dynamical maps $\Phi_{t_{2},t_{1}}$ is said to be P-divisible or CP-divisible if $\Phi_{t_{2},t_{1}}$ is positive or completely positive, respectively, for all $t_{2}>t_{1}$. It turns out that the master equation~\eqref{eq:generalMEform} leads to CP-divisible dynamics if and only if all rates are positive for all times, that is $\gamma_{k}(t)\geq 0 $ which follows from a straightforward extension of the GKSL theorem~\cite{Chruscinski:2012aa}.

The notions of P- and CP-divisibility are intimately related to the notion of quantum non-Markovianity. For our purposes, the relationship between CP-divisibility and quantum non-Markovianity that was first suggested in~\cite{rivas2010entanglement} will suffice (for a more detailed description of the relationship between the two see the reviews~\cite{rivas2014quantum, breuer2016colloquium}). Essentially, the condition \eqref{divisibility} is the quantum
counterpart of the classical Chapman-Kolmogorov equation, and one can make the relationship between classical Markovianity and quantum Markovianity quite precise. 

One important way to detect quantum non-Markovianity in an open quantum system is to measure how the distinguishability of quantum states changes over time. For a non-Markovian process, quantum states should at some times become more distinguishable due to a reverse flow of information from the environment to the open system~\cite{breuer2009measure}. Recall that the trace-norm distance between two quantum states $\rho_{1}$ and $\rho_{2}$ is given by
\begin{equation}
D(\rho_{1},\rho_{2})=\frac{1}{2}\Tr|\rho_{1}-\rho_{2}|,
\end{equation}
where $|A|=\sqrt{A^{\dagger}A}$, and is contractive for any positive and trace-preserving map $\Phi$ \cite{RUSKAI:1994aa} (in particular for any CPTP map), i.e.,
\begin{equation}
D(\Phi(\rho_{1}),\Phi(\rho_{2}))\leq D(\rho_{1},\rho_{2}).
\end{equation}
Suppose that Alice prepares a quantum system in either the state $\rho_{1}$ or $\rho_{2}$. She then hands the system to Bob, and Bob measures the system and decides whether the system was in the state $\rho_{1}$ or $\rho_{2}$. The probability that Bob can successfully identify the state of the system is given by $\frac{1}{2}(1+D(\rho_{1},\rho_{2}))$. Thus we can interpret the trace-norm distance between two quantum states as a measure of distinguishability between the two. 
Because we are interested in the change of distinguishability over time, the following quantity is of particular interest:
\begin{equation}\label{eq:measure1}
\sigma(t,\rho_{1,2}(0))=\frac{d}{dt}D(\rho_{1}(t),\rho_{2}(t)),
\end{equation}
where $\rho_{1,2}(0)$ denotes the initial states. Following~\cite{breuer2009measure}, we will say a process is Markovian if for all pairs of initial states $\sigma(t,\rho_{1,2}(0))\leq 0$ for all times. Therefore, we will say a process is non-Markovian if there exists any pair of initial states $\rho_{1,2}(0)$ and a time $t$ for which $\sigma(t,\rho_{1,2}(0))>0$. However, there is some ambiguity as to whether the process is necessarily Markovian when $\sigma(t,\rho_{1,2}(0))\leq 0$. For instance, an example exists, which we review in the Appendix, where $\sigma(t,\rho_{1,2}(0))\leq 0$ for all $t$ and $\rho_{1,2}(0)$, but the quantum dynamical map associated with the evolution is non-divisible \cite{mazzola2010phenomenological}. 

We note that very recent work~\cite{breuer2018helmstrom} uses an updated witness of information backflow instead of Eq.~\eqref{eq:measure1}. It is calculated by considering the trace of the so-called Helmstrom matrix (essentially a weighted average between two different dynamically evolved initial states), which was shown to also admit an information backflow interpretation. However, for our purposes, the measure given by Eq.~\eqref{eq:measure1} suffices.

\subsection{The post-Markovian master equation}
\label{sec:PMME}
Here we give a brief review of the quantum master equation that is the focus of this paper. Recall that for a quantum system $S$ coupled to a bath $B$ evolving unitarily under the total system-bath Hamiltonian $H_{SB}$, the dynamics of the quantum system have a measurement picture interpretation~\cite{shabani2005completely}. Essentially, the exact system dynamics
\begin{equation}
\rho(t)=\Tr_{B}[U(t)\rho_{SB}(0)U^{\dagger}(t)]
\end{equation}
can be derived by performing a \textit{single} projective measurement on identical ensembles initially prepared in the state $\rho_{SB}(0)$. 
In the Markovian case where the quantum dynamics can be written in the form \eqref{markovlindblad} with $\gamma_{k}\geq0$ $\forall k$, there again exists a measurement interpretation. In the Markovian case the bath functions as a probe coupled to the system while being subjected to a \textit{continuous} series of measurements at an infinitesimal time interval $\tau$. This is related to the well known quantum jump process~\cite{plenio1998quantum}.

The PMME interpolates between these two measurement pictures. In the single measurement picture, exact dynamics are seen as an evolution of the coupled system-bath followed by a single generalized measurement at time $t$. The Markovian dynamics are represented as a series of measurements interrupting the joint evolution after each short time interval $\tau$. The idea is that by relaxing the many-measurements process one is led to a less restricted approximation than the Markovian one.
We skip to the conclusion of the derivation contained in~\cite{shabani2005completely} and give the final form of the PMME:
\begin{equation}\label{eq:pmme}
\frac{d\rho}{dt}=\mathcal{L}\int_{0}^{t}k(t')\exp(\mathcal{L}t')\rho(t-t')dt',
\end{equation}
where $k(t)$ is the memory kernel and $\mathcal{L}$ is the Markovian generator. In the derivation, the memory kernel $k(t)$ is introduced phenomenologically to assign weights to different measurements performed on the bath. While $k(t)$ is left unspecified, it can in principle be determined by an appropriate quantum state tomography experiment.  

Another important feature of the PMME is the dynamical map $\Phi_{t}:\rho(0)\rightarrow\rho(t)$ that governs it. The quantum map corresponding to Eq.~\eqref{eq:pmme} is
\begin{equation}\label{eq:qmap}
\Phi(t):X\rightarrow\sum_{i}\xi_{i}(t)\Tr[L_{i}X]R_{i}
\end{equation}
where the left and right eigenoperators $\{L_{i}\}$ and $\{R_{i}\}$ of the generator $\mathcal{L}$ are known as the damping basis~\cite{briegel1993quantum} of $\mathcal{L}$, and 
\begin{equation}\label{eq:xi}
\xi_{i}(t)=\text{Lap}^{-1}[\frac{1}{s-\lambda_{i}\tilde{k}(s-\lambda_{i})}],
\end{equation}
where $\lambda_{i}$ are the eigenvalues from solving $\mathcal{L}\rho=\lambda\rho$ and the tilde over the kernel represents its Laplace transform. The following condition ensures complete positivity of this dynamical map~\cite{shabani2005completely}:
\begin{equation}\label{eq:cpcondition}
\sum_{k}\xi_{k}(t)L_{k}^{T}\otimes R_{k}\geq 0,
\end{equation}
which results in a condition on the memory kernel $k(t)$. Also, we can expand $\rho(t)$ in the damping basis as 
\begin{equation}\label{eq:rhoexpansion}
\rho(t)=\sum_{i}\xi_{i}(t)\alpha_{i}R_{i},
\end{equation}
 where the $\{\alpha_{i}\}$ can be obtained by expanding $\rho(0)$ in the basis $\{R_{i}\}$. It was shown in the original derivation that for a qubit dephasing with a bath, the solution of this equation indeed interpolates between the exact and Markovian solutions~\cite{shabani2005completely}.

\section{Previous Examples of the non-Markovianity of the PMME}
\label{markov}
The non-Markovianity of the PMME and memory kernel master equations more generally were studied by Mazzola \textit{et al}.~\cite{mazzola2010phenomenological}. They detailed a specific example (re-derived in the Appendix) where the PMME leads to non-divisible quantum dynamics, yet has zero measure for non-Markovianity under Eq.~\eqref{eq:measure1}. They thus included these nondivisible processes which have unidirectional information flow into the class of 
time-dependent Markovian processes. This, of course, does not rule out the possibility of the PMME including non-Markovian effects, given that this analysis was done for a specific example. In fact, in later work by Budini~\cite{budini2014post} it was shown that for a similar example to the one in Ref.~\cite{mazzola2010phenomenological}, if the system Hamiltonian does not commute with the dissipative term in Eq.~\eqref{markovlindblad} then there indeed is backflow of information, and thus the PMME includes non-Markovian effects. More concretely, there is indeed information backflow in the PMME for the generator
\begin{align}
\mathcal{L}_{t}(\rho)&=-i\frac{\Omega}{2}[\sigma_{x},\rho]+\mathcal{C}(\rho) \nonumber \\
&=-i\frac{\Omega}{2}[\sigma_{x},\rho]+\frac{\gamma}{2}([\sigma_{-},\rho\sigma_{+}]+[\sigma_{-}\rho,\sigma_{+}]) ,
\end{align}
with exponential memory kernels. 
The analysis in Ref.~\cite{budini2014post} is particularly useful because it also gives first-principles derivations for the memory kernels used. Also, in 
Ref.~\cite{budini2013embedding} it was  shown that for so-called collisional models, i.e., scenarios where the dissipative term $\mathcal{C}$ can be written as
\begin{equation}
\mathcal{C}(\rho)=\sum_{\alpha}V_{\alpha}\rho V_{\alpha}^{\dagger}-I,
\end{equation}
that using an approximate version of the PMME can lead to non-Markovian effects in the case of qubit dephasing. These two examples give a clear indication for non-Markovian effects present in the PMME. 

We provide an additional perspective in the next section, where we study qubit dephasing with the PMME given by Eq.~\eqref{eq:pmme}, for two choices of memory kernels. We show that the solutions obtained include non-Markovian effects by analyzing both nondivisibility and information backflow.

\begin{figure*}
\centering
\includegraphics[width=\columnwidth]{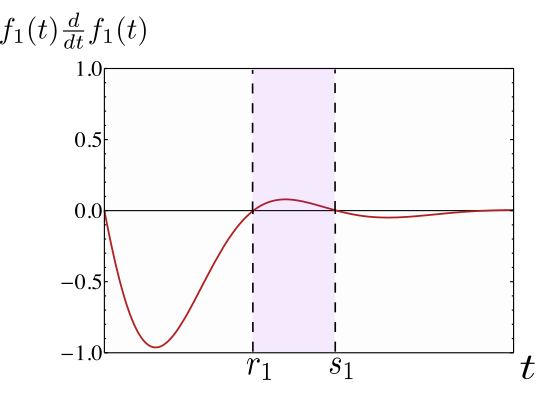}
\includegraphics[width=\columnwidth]{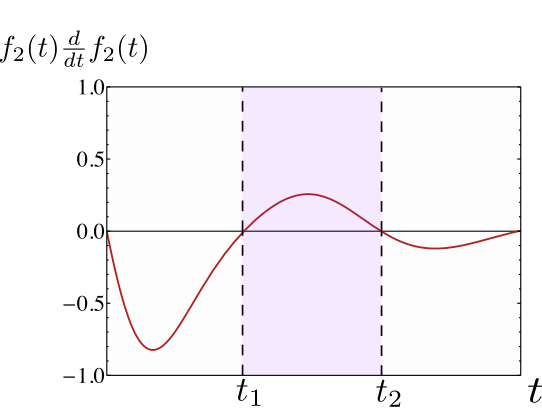}
\caption{The non-Markovianity measure given by Eq.~\eqref{eq:breuermeasure} corresponding to the two different choices of kernels $k_{1}(t)$ (left) and $k_{2}(t)$ (right). Only $f_{1,2}(t)\frac{d}{dt}f_{1,2}(t)$ is plotted since it characterizes the behaviour of $\sigma_{1,2}$ and we can ignore discontinuities induced by the denominator. Note the non-Markovianity regions shown by the shaded regions in both plots. Both solutions exhibit damped oscillations,  and hence an infinite number of non-Markovianity regions, however the plot parameters are chosen so that only one such region is displayed. For $\sigma_{1}$, the zeroes are given by $r_{n}=\frac{2}{\omega}(\pi n-\arctan[\omega/(a+\gamma)])$ and $s_{n}=\frac{n\pi}{\omega}$ where $\omega=\sqrt{4aA-(a+\gamma)^2}$. For $\sigma_{2}$ the zeroes are given by $t_{n}=n\pi/\Omega$ where $\Omega=\sqrt{\mu^{2}+Aa}$. The parameters $A, a$, and $\gamma$ are chosen so that $\omega$ is real or else no non-Markovian effects would be present. The values used for these plots are $A=6, a=1, \gamma=1.1, \text{and }  \mu=\pi$.}
\label{fig:sig1sig2}
\end{figure*}

\section{Non-Markovianity of the PMME via qubit dephasing}
\label{nonmarkov}

Borrowing the example from Ref.~\cite{shabani2005completely}, let us consider the problem of single-qubit dephasing. The GKSL generator is 
\begin{equation}
\mathcal{L}\rho=-\frac{a}{2}[\sigma_{z},[\sigma_{z},\rho]] ,
\end{equation}
where $a>0$. Using the damping basis method~\cite{briegel1993quantum}, we have the following eigenvalues and left and right eigenoperators for the generator $\mathcal{L}$: 
\begin{align}
&\{\lambda_{i}\}_{i}^{3}=\{0,-a,-a,0\},\\
&\{R_{i}\}_{i=0}^{3}=\{L_{i}\}_{i=0}^{3}=\frac{1}{\sqrt{2}}\{I,\sigma_{x},\sigma_{y},\sigma_{z}\}.
\end{align}
It follows immediately from Eq.~\eqref{eq:xi} that $\xi_{0}(t)=\xi_{z}(t)=\text{Lap}^{-1}[1/s]=1$ and $\xi_{x}(t)=\xi_{y}(t)\equiv f(t)$ where $f(t)$ can be given explicitly once we have chosen a kernel. From Eq.~\eqref{eq:rhoexpansion} we have
\begin{equation}\label{eq:rhoexp}
\rho(t)=\frac{1}{2}\big(I+f(t)\alpha_{x}(0)\sigma_{x}+f(t)\alpha_{y}(0)\sigma_{y}+\alpha_{z}(0)\sigma_{z}\big).
\end{equation}
 To proceed we must make a choice for the kernel function. Consider the following simple memory kernels~\cite{shabani2005completely}:
\begin{align}
&k_{1}(t)=Ae^{-\gamma t},\\
&k_{2}(t)=Ae^{-(\gamma-a)t}[\cos(\mu t)-\frac{\gamma}{\mu}\sin(\mu t)].
\end{align}
The rather specific form of $k_{2}(t)$ is because the associated solution is known to lead to damped oscillatory dynamics with a non-zero asymptotic coherence, which is a feature of the exact solution for a single qubit dephasing in the presence of a boson bath. 

We are now in a position to analyze these solutions with the non-Markovianity measure given by Eq.~\eqref{eq:measure1}. To do so, we first expand two evolved density matrices $\rho_{u}(t)$ and $\rho_{v}(t)$ as in Eq.~\eqref{eq:rhoexp} and take their difference to obtain the matrix
\begin{align}
&A=\rho_{u}-\rho_{v} \nonumber \\
&=
\begin{pmatrix}
\Delta^{(u,v)}_{z} & (\Delta^{(u,v)}_{x}-i\Delta^{(u,v)}_{y})f(t) \\
(\Delta^{(u,v)}_{x}+i\Delta^{(u,v)}_{y})f(t) & -\Delta^{(u,v)}_{z}
\end{pmatrix}
\end{align}
where we use $\Delta^{(u,v)}_{i}=\alpha^{(u)}_{i}(0)-\alpha^{(v)}_{i}(0)$, $i=x,y,z$ to denote the difference of Bloch vector coefficients between $\rho_{u}(0)$ and $\rho_{v}(0)$. 
Computing the trace of $\frac{1}{2}|\rho_{u}-\rho_{v}|=\frac{1}{2}\sqrt{A^{\dagger}A}=\frac{1}{2}\sqrt{A^{2}}$ gives 
\begin{align}
\frac{1}{2}&\Tr|\rho_{u}-\rho_{v}| \nonumber \\
&=\sqrt{f(t)^{2}[(\Delta^{(u,v)}_{x})^{2}+(\Delta^{(u,v)}_{y})^{2}]+(\Delta^{(u,v)}_{z})^{2}}.
\end{align} 
Finally, by taking the time derivative derivative of this quantity, we arrive at 
\begin{equation}\label{eq:breuermeasure}
\sigma_{1,2}(t,\rho_{u,v}(0))=\frac{(\Delta_{x}^{2}+\Delta_{y}^{2})f_{1,2}(t)\frac{d}{dt}f_{1,2}(t)}{\sqrt{(\Delta_{x}^{2}+\Delta_{y}^{2})(f_{1,2}(t))^{2}+\Delta_{z}^{2}}},
\end{equation}

where
\begin{align}
&f_{1}(t)=e^{-t(a+\gamma)/2}[\cos(\omega t)+\sin(\omega t)(a+\gamma)/2\omega]\label{eq:f1},\\
&f_{2}(t)=1-\frac{Aa}{\gamma^{2}+\Omega^{2}}[1-e^{-\gamma t}(\cos\Omega t+\frac{\gamma}{\Omega}\sin\Omega t)],\label{eq:f2}
\end{align}
$\omega=\sqrt{4aA-(\gamma+a)^{2}}/2$, $\Omega=\sqrt{\mu^{2}+Aa}$, and $\Delta^{(u,v)}_{i}=\Delta_{i}$ to simplify notation. Note that the condition for complete positivity [Eq.~\eqref{eq:cpcondition}] results in $|f_{1,2}(t)|\leq 1$~\cite{shabani2005completely}, which imposes restrictions on the allowed values of the various parameters appearing here, but we are mainly interested in the damped oscillatory nature of the solutions. We choose the various parameters such that $\omega$ is real, which ensures that $f_{1}(t)$ is oscillatory. To demonstrate non-Markovianity from information backflow from the bath to the system it is sufficient to show that $f_{1,2}(t)\frac{d}{dt}f_{1,2}(t)$ can become positive. This is illustrated in Fig.~\ref{fig:sig1sig2}.

We can go further by explicitly checking whether the dynamical map $\Phi_{t}$ associated with this evolution is CP-divisible or not. Upon inspection of Eq.~\eqref{eq:rhoexp}, we see that for any $t_{1}>0$, $\Phi_{t_{1}}$ must have the following form:
\begin{equation}
\rho(t_{1})=\Phi_{t_{1}}(\rho(0))=a_{1}\rho(0)+b_{1}\sigma_z\rho(0) \sigma_z
\end{equation}
where $a_{1}>0$ and $b_{1}>0$ satisfy $a_{1}+b_{1}=1$. By comparing the Bloch vector expansions of both $\rho(0)$ and $\rho(t)$ we can solve for the coefficients $a_{1},b_{1}$:
\begin{align}\label{eq:a1b1}
a_{1}=\frac{1+f(t_{1})}{2},\quad
 b_{1}=\frac{1-f(t_{1})}{2}.
\end{align}
Now consider the map $\Phi_{t_{2},t_{1}}$ for $t_{2}>t_{1}$. This map acting on $\rho(t_{1})$ yields
\begin{align}
\rho(t_{2})&=a_{2}\rho(t_{1})+b_{2}\sigma_z\rho(t_{1})\sigma_z \nonumber \\
&=(a_{2}a_{1}+b_{2}b_{1})\rho(0)+(a_{2}b_{1}+b_{2}a_{1})\sigma_z\rho(0)\sigma_z.
\end{align}
Therefore 
\begin{align}
a_{2}a_{1}+b_{2}b_{1}=\frac{1+f(t_{2})}{2},\quad
 a_{2}b_{1}+b_{2}a_{1}=\frac{1-f(t_{2})}{2},
\end{align}
and by plugging this solution into Eq.~\eqref{eq:a1b1} and solving for $a_{2},b_{2}$ we arrive at
\begin{align}\label{eq:a2b2}
a_{2}=\frac{1}{2}\big(1+\frac{f(t_{2})}{f(t_{1})}\big), \quad
b_{2}=\frac{1}{2}\big(1-\frac{f(t_{2})}{f(t_{1})}\big).
\end{align}
Now consider the case where $t_{2}=t_{1}+h$ where $h>0$ is small. Then we can rewrite $f(t_{2})=f(t_{1}+h)\approx f(t_{1})+\frac{d}{dt}f(t_{1})h$, and Eq.~\eqref{eq:a2b2} becomes 
\begin{align}
a_{2}=1+\frac{1}{2}\frac{\frac{d}{dt}f(t_{1})}{f(t_{1})}h , \quad
b_{2}=-\frac{1}{2}\frac{\frac{d}{dt}f(t_{1})}{f(t_{1})}h.
\end{align}
From this we see that if ${\frac{d}{dt}[f(t_{1})]}/{f(t_{1})}>0$, then $\Phi_{t_{2},t_{1}}$ is not a valid Kraus map. Indeed, we can infer from Fig.~\ref{fig:sig1sig2} that there exist time intervals for which this inequality holds for both choices of kernel. Hence we do not have CP-divisible dynamics, as expected from our analysis of $\sigma_{1,2}$.

Furthermore, we can write down a quantum master equation of the form given by Eq.~\eqref{eq:generalMEform} for these dynamics generated by the PMME. We follow the method given in Ref.~\cite{pang2017abrupt} for how to do this in general when the solution to the system evolution is known. From Eq.~\eqref{eq:rhoexp}, we see that
\begin{equation}
\dot{\rho}(t)=\frac{1}{2}\frac{d}{dt}f(t)(\alpha_{x}(0)\sigma_{x}+\alpha_{y}(0)\sigma_{y}).
\end{equation}
If we denote the vector of Bloch coefficients of $\rho(t)$ by $\vec{\alpha}(t)$, then it can be easily checked that
\begin{equation}
\frac{d}{dt}\vec{\alpha}(t)=Q\vec{\alpha}(t),
\end{equation}
where 
\begin{equation}
Q=
\begin{pmatrix}
0 & 0 & 0 &0 \\
0 & \frac{d}{dt}[f(t)]/f(t) & 0 & 0 \\
0 & 0 & \frac{d}{dt}[f(t)]/f(t) & 0 \\
0 & 0 & 0 & 0
\end{pmatrix}
.
\end{equation}
Now we can find the superoperator corresponding to $Q$. Note that there are 16 basis elements $\sigma_{i}[\cdot]\sigma_{j}$ for the superoperator which we denote by $s_{ij}$. The basis elements
\begin{equation}
s_{zz}=
\begin{pmatrix}
1 & 0 & 0 & 0 \\
0 & -1 & 0 & 0 \\
0 & 0 & -1 & 0 \\
0 & 0 & 0 & 1
\end{pmatrix}
, \hspace{2mm} s_{00}=
\begin{pmatrix}
1 & 0 & 0 & 0 \\
0 & 1 & 0 & 0 \\
0 & 0 & 1 & 0 \\
0 & 0 & 0 & 1 
\end{pmatrix}
\end{equation}
allow us to write $Q= -\frac{1}{2}\frac{d}{dt}[f(t)]/f(t)(s_{zz}-s_{00})$. Therefore the quantum master equation of the form given by Eq.~\eqref{eq:generalMEform} for these dynamics is 

\begin{equation}
\dot{\rho}=\frac{\gamma(t)}{2}(\sigma_{z}\rho\sigma_{z} - \rho),
\end{equation}
where $\gamma(t)=-\frac{d}{dt}[f(t)]/f(t)$ (note the negative sign). As expected, there exist time intervals where $\gamma(t)<0$ for both memory kernels (as can be inferred from Fig.~\ref{fig:sig1sig2}), which is consistent with the extended GKSL theorem~\cite{Chruscinski:2012aa} and our non-Markovian analysis thus far. 

\section{Which kernels give rise to CP-divisible dynamics?}\label{Cpkerncondition}
A natural question to ask is what are the classes of kernels $k(t)$ that give rise to CP-divisible dynamics for the PMME? Given a general Markovian generator $\mathcal{L}$ (and hence the eigenoperators $\{L_{i}, R_{i}\}$ and eigenvalues $\{\lambda_{i}\}$), the condition on the kernel $k(t)$ for the associated dynamics to be CP divisible can be derived from Eq.~\eqref{eq:cpcondition}. First we derive the quantum map which maps $\rho(t)$ to $\rho(t+dt)$. Multiplying both sides of Eq.~\eqref{eq:qmap} by $L_{j}$ and taking the trace, we have
$\Tr[L_{j}\rho(t)]=\sum_{i}\xi_{i}(t)\Tr[L_{i}\rho(0)]\Tr[L_{j}R_{i}] 
=\xi_{j}(t)\Tr[L_{j}\rho(0)]$,
where in the second equality we used the fact that $\Tr[L_{j}R_{i}]=\delta_{ij}$. Therefore
\begin{equation}\label{eq:cpderiv}
\Tr[L_{j}\rho(0)]=\frac{\Tr[L_{j}\rho(t)]}{\xi_{j}(t)}
\end{equation}
for all $t$ where $\xi_{j}(t)\neq 0$. Note that from Eq.~\eqref{eq:qmap} we also have 
\begin{equation}
\rho(t+dt)=\sum_{i}\xi_{i}(t+dt)\Tr[L_{i}\rho(0)]R_{i},
\end{equation}
so combining this with Eq.~\eqref{eq:cpderiv} gives
\begin{equation}\label{eq:ttodt}
\rho(t+dt) = \sum_{i}\frac{\xi_{i}(t+dt)}{\xi_{i}(t)}\Tr[L_{i}\rho(t)]R_{i}.
\end{equation}
For CP-divisible dynamics, this map must be completely positive. By applying Eq.~\eqref{eq:cpcondition} to Eq.~\eqref{eq:ttodt}, we arrive at the condition for CP-divisible dynamics:
\begin{equation}\label{eq:cpkern}
\sum_{i}\frac{\xi_{i}(t+dt)}{\xi_{i}(t)}L_{i}^{T}\otimes R_{i} \geq 0.
\end{equation}
Because the functions $\xi_{i}$ are given in terms of the memory kernel $k(t)$ through Eq.~\eqref{eq:xi}, this inequality gives a condition on $k(t)$ that can be checked to verify that the given kernel produces CP divisible dynamics. Given our analysis so far, we expect Eq.~\eqref{eq:cpkern} to be violated for the qubit dephasing example studied in Section \ref{nonmarkov}. The left-hand side of Eq.~\eqref{eq:cpkern} for this example becomes
\begin{equation}
\frac{1}{2}\big( I+(1+\frac{\frac{d}{dt}f(t)}{f(t)})\big(\sigma_{x}\otimes\sigma_{x}+\sigma_{y}^{T}\otimes\sigma_{y}\big)+\sigma_{z}\otimes\sigma_{z}\big).
\end{equation}
The eigenvectors of this operator are given by the Bell basis, and from a straightforward calculation we see that the eigenvalues are 
\begin{equation}
\big\{\lambda_{i}\big\}_{i=1}^{4}=\{0,0,\gamma(t),2+\gamma(t)\}
\end{equation}
where $\gamma(t)=-\frac{d}{dt}[f(t)]/f(t)$. Since $\lambda_{3}<0$ when $\gamma(t)<0$ (which corresponds precisely to the non-Markovianity region in Fig.~\ref{fig:sig1sig2}), the CP-divisibility condition given by Eq.~\eqref{eq:cpkern} is violated for both memory kernels, as expected.
\section{Conclusions and Future Work}\label{conclusion}
In conclusion, we have shown through a simple example that the post Markovian master equation (PMME) can describe non-Markovian effects. We did this by analyzing the change of distinguishability of quantum states, and by checking the divisibility of the associated quantum maps. This complements the much more general non-Markovian analysis of the PMME given in Ref.~\cite{budini2014post}. 

Quantum non-Markovianity is a good metric for judging whether solutions obtained from the PMME really do interpolate between the Markovian master equation and the exact (Nakajima-Zwanzig) equation. Ultimately, being able to describe as many different physical scenarios as possible, with different bath characterizations (memory kernels) and an easily solvable master equation is an important step towards modeling of open system quantum dynamics that is both rich and tractable. Future work should will this analysis in more physically motivated scenarios. 

\acknowledgments
We thank Shengshi Pang and Yi-Hsiang Chen for helpful discussions. TAB and CS were funded in part by NSF Grant No. CCF-1421078, NSF Grant No. QIS-1719778, and by an IBM Einstein Fellowship at the Institute for Advanced Study. This material is based upon work supported by the Intelligence Advanced Research Projects Activity (IARPA) through the Army Research Office (ARO) under Contract No. W911NF-17-C-0050. Any opinions, findings and conclusions or recommendations expressed in this material are those of the author(s) and do not necessarily reflect the views of the Intelligence Advanced Research Projects Activity (IARPA) and the Army Research Office (ARO).

\appendix*
\section{An example with no information backflow but with a non-divisible map}
We review an example due to Ref.~\cite{mazzola2010phenomenological} that illustrates the subtlety of defining non-Markovianity purely via information backflow from the environment.

Consider the dynamics of a spin-$1/2$ particle interacting with a bosonic reservoir at temperature $T$. The Markovian generator associated with this process is 
\begin{align}
 \mathcal{L}\rho=&\frac{\gamma_{0}}{2}(N+1)(2\sigma_{-}\rho\sigma_{+}-\sigma_{+}\sigma_{-}\rho-\rho\sigma_{+}\sigma_{-}) \nonumber \\
&+\frac{\gamma_{0}}{2}N(2\sigma_{+}\rho\sigma_{-}-\sigma_{-}\sigma_{+}\rho-\rho\sigma_{-}\sigma_{+}),
\end{align}
where $\gamma_{0}$ is the dissipation rate, $N$ is the mean number of excitations of the reservoir, and $\sigma_{\pm}$ are the raising and lowering operators. Using the kernel function $k(t)=\gamma e^{-\gamma t}$ we can solve the PMME by following the procedure given in Sec.~\ref{sec:PMME}. The result is:
\begin{align}
&\rho(t)=\frac{N}{1+2N}
\begin{pmatrix}
\frac{1+N}{N} & 0\\
0 & 1
\end{pmatrix}
+\xi(A,B,t)\bigg[b^{*}
\begin{pmatrix} \nonumber
0 & 0 \\
1 & 0
\end{pmatrix}
\\
&+b
\begin{pmatrix}
0 & 1 \\
0 & 0
\end{pmatrix}
+(1-a-\frac{N}{1+2N})
\begin{pmatrix}
-1 & 0 \\
0 & 1
\end{pmatrix}
\bigg]
\end{align}
where the initial state is $\rho(0)=\begin{pmatrix}a & b\\ b^{*} & 1-a\end{pmatrix}$, where $A=(1+2N)\gamma\gamma_{0}$, $B=\gamma+(1+2N)\gamma_{0}$, and where 
\begin{align}
\xi(A,B,t)= &e^{-\frac{Bt}{2}}\big[\cosh\big(\frac{t}{2}\sqrt{B^{2}-4A}\big) \nonumber \\
&+\frac{B}{\sqrt{B^{2}-4A}}\sinh\big(\frac{t}{2}\sqrt{B^{2}-4A}\big)\big].
\end{align}
 We can now compute the non-Markovianity measure given by Eq.~\eqref{eq:measure1}. For initial  states $\rho_{1}(0)=\begin{pmatrix}a & 0 \\ 0 & 1-a \end{pmatrix}$ and $\rho_{2}(0)=\begin{pmatrix}c & 0 \\ 0 & 1-c \end{pmatrix}$ we have
\begin{equation}
\sigma(t,\rho_{1,2}(0))=-\frac{4Ae^{-\frac{Bt}{2}}|a-c|\sinh(\frac{t}{2}\sqrt{B^{2}-4A})}{\sqrt{B^{2}-4A}}
\end{equation}
which is always negative since $B^{2}-4A=(\gamma+(1+2N)\gamma_{0})^{2}$ is a perfect square. It is straightforward to verify that this is in fact true for all initial states $\rho_{1}$ and $\rho_{2}$. Hence the solution contains no information backflow from the bath to the system, and so one might be tempted to conclude that it is Markovian. However, as shown in Ref.~\cite{mazzola2010phenomenological}, the associated time-local quantum master equation of the form given by Eq.~\eqref{eq:generalMEform} has a negative rate coefficient, i.e., $\gamma_{3}\leq 0$. Therefore the dynamical map $\Phi$ corresponding to the master equation is nondivisible, in contrast to Markovian processes. This example highlights the subtlety in defining quantum non-Markovianity via the concept of information backflow.

\bibliography{pmmenonmarkovianity,refs}

\end{document}